\documentclass{ar}
\usepackage[numbers]{natbib}
\usepackage{makecell}
\usepackage[table]{xcolor}
\jname{Xxxx. Xxx. Xxx. Xxx.}
\jvol{AA}
\jyear{YYYY}
\doi{10.1146/((please add article doi))}

\begin{document}

\markboth{Schoop et al.}{Topology and Square Nets}

\title{Topological Semimetals in Square-Net Materials}

\author{Sebastian Klemenz,$^1$, Shiming Lei$^1$ and Leslie M. Schoop,$^1$
\affil{$^1$Department of Chemistry,Princeton University, Princeton, NJ USA, 08540; email: lschoop@princeton.edu}
\affil{$^*$All authors contributed equally}}

\begin{abstract}
Many materials crystallize in structure types that feature a square-net of atoms. While these compounds can exhibit many different properties, some members of this family are topological materials. Within the square-net-based topological materials, the observed properties are rich, ranging for example from nodal-line semimetals to a bulk half-integer quantum Hall effect. Hence, the potential for guided design of topological properties is enormous. Here we provide an overview of the crystallographic and electronic properties of these phases and show how they are linked, with the goal of understanding which square-net materials can be topological, and to predict additional examples. We close the review by discussing the experimentally observed electronic properties in this family. 
\end{abstract}

\maketitle

\tableofcontents

\section{INTRODUCTION}
Square nets of atoms, which can appear in a large variety of crystal structures, have been linked to many exciting physical properties. The best known example is probably the square net in cuprate superconductors, often linked to the extremely high superconducting transition temperatures in those materials.  \cite{cava2000oxide,shaked1994crystal}. Similarly, in iron-based superconductors, the iron atoms are arranged in a square net  \cite{mazin2010superconductivity,paglione2010high}. Square nets are relatively easy to model theoretically, and so have served as the basis for a variety of predictions of materials properties \cite{kramers1941statistics,malliakas2005square}. In recent times, the square net has been linked to a very different kind of material property. Many intermetallic compounds that are isostructural or structurally related to iron-based superconductors  have been shown to be topological semimetals (TSMs) \cite{park2011anisotropic,xu2015two,schoop2015dirac,borisenko2015time,masuda2016quantum,hu2016evidence,huang2017nontrivial,bradlyn2017topological,liu2017magnetic,schoop2017tunable,kealhofer2018observation}. 

\subsection{A brief description of topological semimetals}

\begin{figure}[h]
\includegraphics[width=4in]{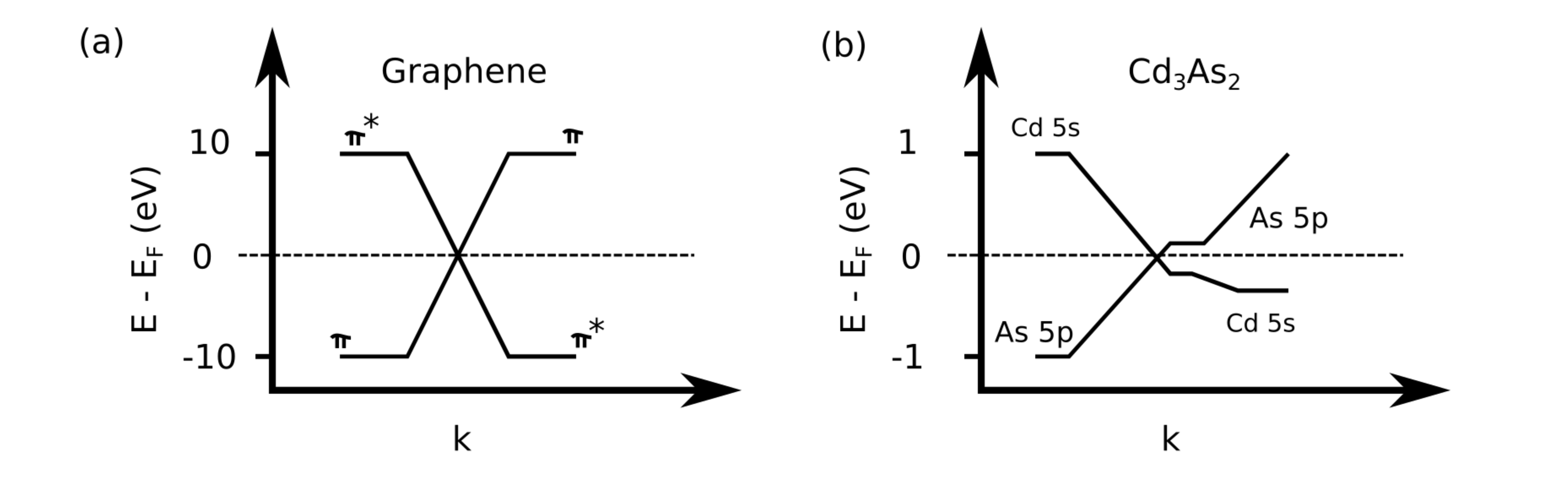}
\caption{Schematic band structure of (a) graphene and (b) Cd$_3$As$_2$. Note the different energy scales in (a) and (b).}
\label{CvsCd3As2}
\end{figure}

TSMs are materials that are electronically analogous to graphene. Their complexity can go beyond that of graphene, however, and the term ``topological semimetal" also includes materials in which the electrons do not behave as typical Dirac fermions, but rather can be described by different Hamiltonians, e.g. as Weyl fermions \cite{vafek2014dirac,xu2015discovery}. For an in-depth description of TSMs and the chemical principles behind them, we like to refer readers to a recent review article \cite{schoop2018chemical}. The electronic structure of a TSM consists of two linearly dispersed bands that cross each other at the so-called Dirac (or Weyl) point, which is ideally located at the Fermi level. In graphene, the linearly dispersing bands extend over a very wide energy range (more than 15 eV), and no other bands interfere in an energy range of roughly 10 eV \cite{wallace1947band}. \textbf{Figure \ref{CvsCd3As2} (a)} sketches the band structure of graphene qualitatively. The $\pi$ and $\pi^\ast$ bands are inverted beyond the Dirac point. The electronic structure of graphene is a result of the half filled carbon \textit{p$_z$}-orbitals, which are arranged in a honeycomb (see \cite{schoop2018chemical} for details). The first 3D analogs of graphene, Cd$_3$As$_2$ and Na$_3$Bi, exhibit a much smaller region of linearly dispersed bands, far less than 1 eV (\textbf{Figure \ref{CvsCd3As2} (b)}). The reason lies in the different origin of the band inversion. While in graphene, the delocalized $\pi$-\textit{bonds} are responsible for the band inversion \cite{schoop2018chemical}, which in turn cause a very large ``band inversion energy", in Na$_3$Bi (or Cd$_3$As$_2$), the low-lying \textit{6s} (or \textit{5s}) \textit{atomic orbital} of Bi (Cd) causes the band inversion \cite{wang2012dirac, wang2013three}. For this reason, topological materials that are similar to graphene have previously been called ``molecular" topological materials, while those that are based on an energetically low \textit{s} orbitals have been called ``atomic" topological materials \cite{wang2016hourglass}. Band inversions caused by a low-lying orbital are rarely very large, i.e the linear bands don't extend far in energy below the crossing point.  If a band inversion is driven by bonding between atoms, however, the linear regime can be extremely large, as exemplified by graphene.  
Whether chemical bonding can drive a band inversion depends on a material's crystal structure and electron count. Structural motifs can often be linked to electronic structure features. In recent years, many materials with the square net motif have been discovered to be TSMs, with very large band inversion energies. While lower than in graphene, the band inversion energy is much larger than in Cd$_3$As$_2$ or Na$_3$Bi. Therefore, we explore the hypothesis that square nets might be a structural motif that drive band inversions as a result of chemical bonding. This motivated us to review the square-net-based TSMs and what is known about their basic electronic structures.

\subsection{A brief history of square-net topological semimetals}

\begin{figure}[h]
\includegraphics[width=5in]{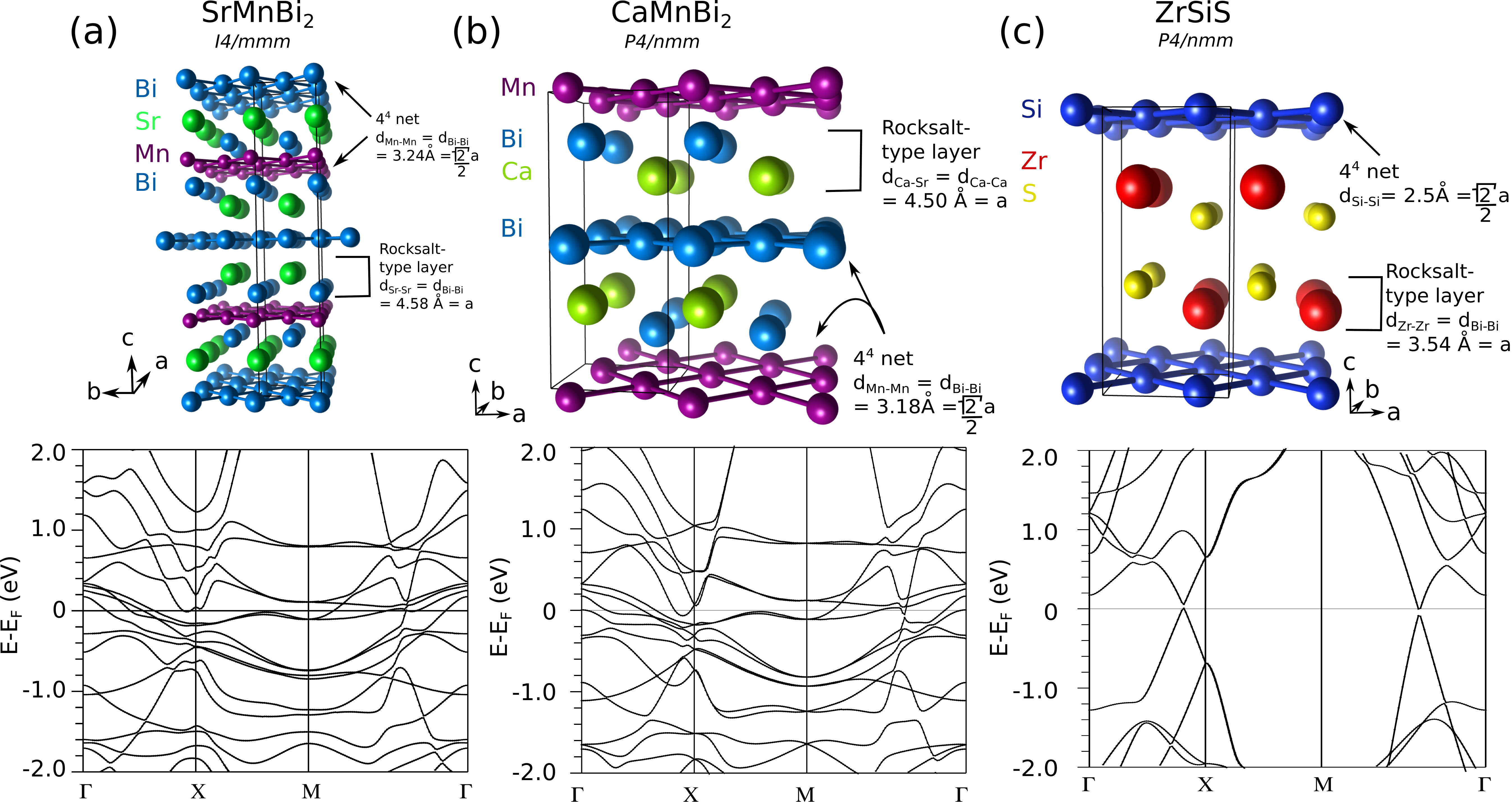}
\caption{(a) Crystal and electronic structure of SrMnBi$_2$. The unit cell information for the band structure calculation is taken from \cite{cordier1977darstellung}. Note that the band structure is plotted for the primitive BZ to allow for an easier comparison with the materials shown in (b) and (c). (b) Crystal and electronic structure of CaMnBi$_2$. The unit cell information for the band structure calculation is taken from ref. \cite{brechtel1980darstellung}. (c) Crystal and electronic structure of ZrSiS. The band structure has been reproduced from \cite{schoop2015dirac}. Note: the band structures in (a) and (b) were calculated based on a paramagnetic model for the convenience of comparison with paramagnetic ZrSiS in (c). For the closely related band structure of antiferromagnetic SrMnBi$_2$ and CaMnBi$_2$, see refs. \cite{park2011anisotropic,lee2013anisotropic}. All calculations account for SOC.}
\label{strucs}
\end{figure}

 The interest in square-net materials as TSMs began with the discovery of anisotropic Dirac fermions in SrMnBi$_2$ \cite{park2011anisotropic}. In this layered tetragonal material (CaSmP$_2$-type, space group \textit{I4/mmm}), Mn and Bi each form square nets with relatively small atomic distances (the distance is $\sqrt{2}/2\cdot a$, with $a$ being the in-plane lattice constant). These densely packed square nets are commonly referred to as the $4^4$-net in the crystallography literature \cite{hoffmann1987chemistry}. In  the SrMnBi$_2$  structure, they are stacked with puckered rock salt-like SrBi slabs (see \textbf{Figure \ref{strucs} (a)} ). As can be seen in \textbf{Figure \ref{strucs} (a)}, the structure of SrMnBi$_2$ is rather complex, resulting in a complex band structure. The linear bands appear quite hidden in the band structure. 
 
 The closely related compound CaMnBi$_2$ crystallizes in a simpler crystal structure in space group \textit{P4/nmm}, where the layers are stacked differently, resulting in a unit cell that contains half the amount of atoms (see \textbf{Figure \ref{strucs} (b)}). The structure type of CaMnBi$_2$ is commonly referred to as the HfCuSi$_2$ structure type and it is isostructural to the '1111'-type Fe-based superconductors \cite{schoop2013superconductivity} . Even though this structure is simpler, the electronic structure remains complex and the analogy to graphene is not obvious (see \textbf{Figure \ref{strucs} (b)}). Roughly five years after the discovery of Dirac fermions in SrMnBi$_2$, it was found that the electronic structure of materials in the PbFCl structural family (such as ZrSiS) also show linearly dispersed bands. Here, the bands cross at the Fermi level and they can be linear over an energy range of up to 4 eV \cite{xu2015two,schoop2015dirac}. These materials exhibit much cleaner electronic structures, without bands interfering with the linear bands at the Fermi level (\textbf{Figure \ref{strucs} (c)}). The PbFCl structural family is closely related to the HfCuSi$_2$ structure, which can be viewed as a stuffed version of PbFCl. 
 
The analogy of the electronic structure of square-net materials to graphene can be clearly seen in ZrSiS. As in graphene, the linearly dispersed bands cross exactly at the Fermi level and no other bands interfere in the bulk electronic structure. Nevertheless, there are a few important differences. In ZrSiS, the bands cross more than once within the Brillouin zone (BZ). If plotted in 3D, these crossings result in a diamond-shaped Fermi surface (FS), rather than a dot-shaped FS as in graphene. Square-net materials are thus usually nodal-line semimetals (NLSMs), which are TSMs that have a 1D FS rather than a 0D one. Another important aspect is that spin-orbit-coupling (SOC) is larger in the square-net materials of interest here. SOC creates a gap at the Dirac crossing points, even graphene theoretically has a small gap, but it is too small to be experimentally resolved \cite{gmitra2009band}. In ZrSiS, the gap is still very small (smaller than 30 meV), but it is measurable \cite{schilling2017flat}. In the Bi-based square-net materials SOC is large and forms another cause for the more complicated electronic structure.
 
 Besides the three structure types mentioned above, there are many additional related structures that exhibit square nets \cite{benavides2017casting,phelan2011adventures}. Many of them are reported to be topologically non-trivial or to exhibit unusual physical properties \cite{bradlyn2017topological,benavides2017casting}. For the sake of brevity, this review will focus on PbFCl- and HfCuSi$_2$-type materials, based on which many of the general characteristics of square-net topological materials can be understood. 
  






\section{THE ELECTRONIC STRUCTURE OF A SQUARE NET}

\subsection{Deriving a band structure for a square net}

\begin{figure}[h]
\includegraphics[width=3.5in]{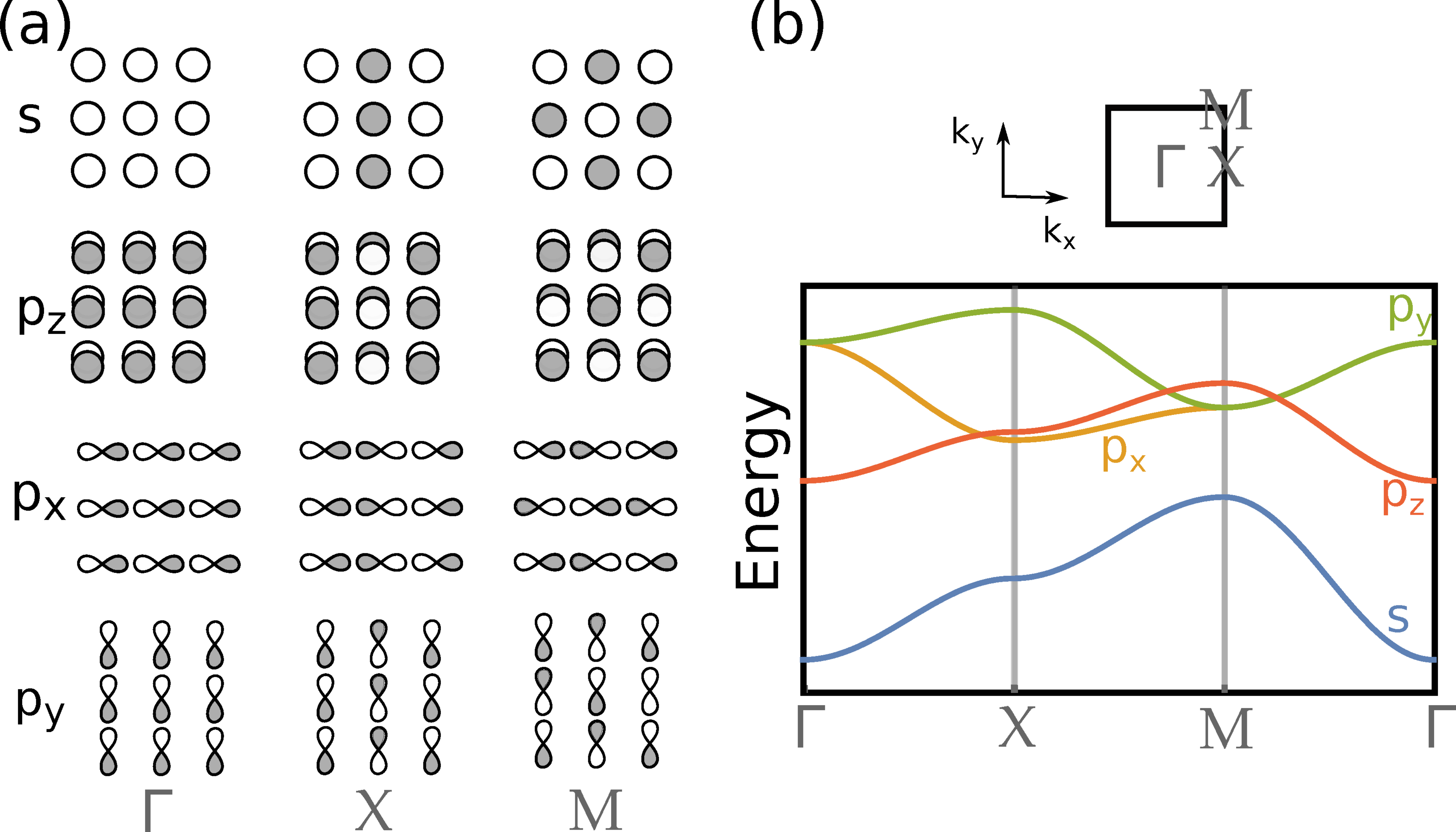}
\caption{(a) Phases or \textit{s}- and \textit{p}-orbitals in a square net. White shades indicate a positive sign and grey shades a negative sign of the wave function. (b) resulting electronic structure}
\label{orbitals}
\end{figure}

\begin{figure}[h]
\includegraphics[width=3.5in]{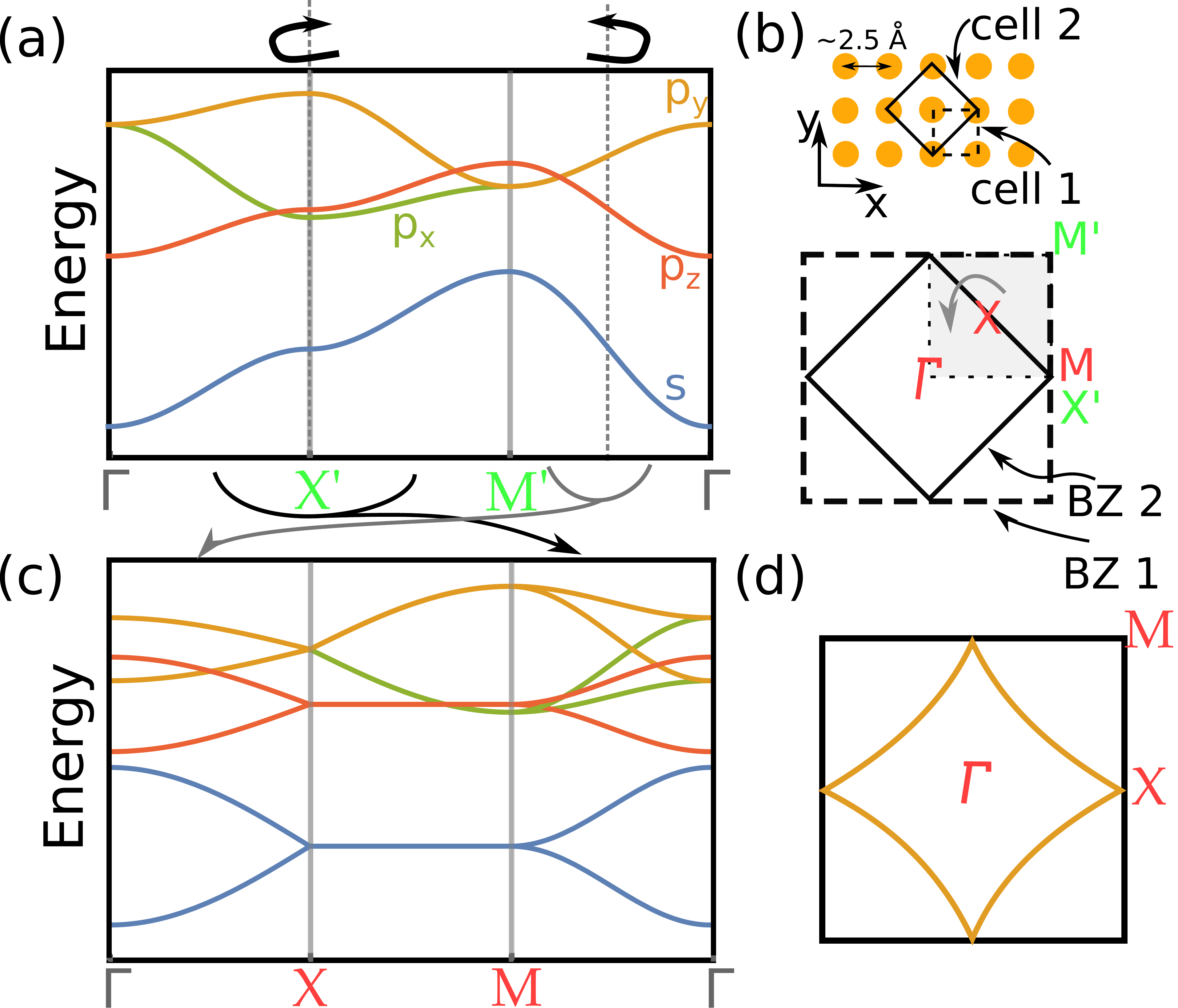}
\caption{(a) Band structure of a square net based on the orbitals drawn in \textbf{Figure \ref{orbitals}}. (b) The 4$^4$ square net folds the BZ. (c) Folded band structure.(d) Fermi surface for a hypothetical material where the Fermi level crosses the \textit{p$_x$} and \textit{p$_y$} bands.}
\label{sqbands}
\end{figure}

In his paper entitled ``How chemistry and physics meet in the solid state" from 1987, \textit{Roald Hoffmann} described an intuitive way for chemists to derive simple band structures \cite{hoffmann1987chemistry}. The Bloch function describes how often an orbital should switch phases at high-symmetry points in the BZ.  Combined with a chemist's intuition of how strong or weak the orbital overlap will be, we can sketch a band structure without using any type of computational method. It is therefore possible to draw a simple band structure of \textit{s}- and \textit{p}-orbitals that are arranged in a square-net by hand. \textit{Tremel} and \textit{Hoffmann} used this method (combined with density functional theory (DFT) calculations) to investigate the electronic structure of square-net materials \cite{tremel1987square}. We will briefly review their work in this section. \textbf{Figure \ref{orbitals} (a)} shows a drawing of \textit{s}- and \textit{p}-orbitals in a square-net arrangement and their phases at different high-symmetry points of a square BZ. At the $\Gamma$ point, all orbitals have the same phase. Therefore, all \textit{s}-orbitals will have bonding interactions resulting in an \textit{s}-band that has an energy minimum at $\Gamma$. At the X point, the phase of the orbitals will switch along $k_x$, but not $k_y$, while at the M point the phase will switch along both directions. Thus, the \textit{s}-band will be highest in energy at the M point. The \textit{p}-bands can be drawn in a similar way. Note that the \textit{p}-bands will generally be located at higher energies than the \textit{s}-bands and the $\pi$-bonding causes a smaller band dispersion than $\sigma$-bonding. \textbf{Figure \ref{orbitals}(b)} shows a sketch of the resulting band structure for \textit{s}- and \textit{p}-orbitals in a square net. It is drawn based on a tight-binding (TB) model with selected parameters that match the chemical intuition and do not actually represent results from real atomic distances. For a complete discussion of how to derive such a band structure, please see refs. \cite{tremel1987square,hoffmann1987chemistry}. Note, that the magnitude of the band dispersion depends on the atomic distance. In ZrSiS, for example, the Si-Si distance is 2.5$\,\mathrm{\AA}$, as indicated in \textbf{Figure \ref{sqbands}(b)}. Only if the distances of the atoms composing the square nets are sufficiently small, the bands will show dispersion, but for large distances, they will be flat.

The band structure of the square net shows some band crossings but there is no obvious linear dispersion, which would indicate that the square net is a good structural motif for TSMs. In order to understand why so many square-net materials show very steep linearly dispersed bands, we first need to look more closely at their crystal structure. As an example, let us consider the PbFCl crystal structure. Compounds in this structure type have the general composition \textit{MXZ}. The \textit{M} and the \textit{Z} element form a puckered rock salt-like layer. One could also view the rock salt-like layer as four square nets that are stacked on top of each other in a staggered fashion. The atomic distances in these square nets are relatively large (the size of the $a$ lattice constant of the unit cell), and usually the in-plane bonding in these layers can be negated. The \textit{X} element forms a more densely packed square net (the $4^4$ net), where the bond distance between the atoms is $\sqrt{2}/2\cdot a$. Thus, the $4^4$ net is packed twice as dense and the unit cell contains two \textit{X} atoms within one layer (see \textbf{Figure \ref{strucs}(c)}). If we go back to the 2D unit cell for the square net, we have to enlarge the unit cell to account for the two \textit{X} atoms of the $4^4$ net (see \textbf{Figure \ref{sqbands}(b)}). This larger unit cell in real space will cause a reduction (or folding) of the BZ in reciprocal space, as indicated in \textbf{Figure \ref{sqbands}(b)}. The folding process can be pictured as if a square-shaped piece of paper is folded in a way in which all four corners are folded into the middle, resulting in a smaller square. This way, the M point of the larger BZ (denoted M' in \textbf{Figure \ref{sqbands}}) folds onto the $\Gamma$ point of the smaller BZ (see \textbf{Figure \ref{sqbands}(b)}). \textbf{Figure \ref{sqbands}(a)} indicates how this folding will affect the band structure. The resulting folded band structure is shown in \textbf{Figure \ref{sqbands}(c)}. Here, linear band crossings are visible between \textit{p}-bands along $\Gamma$X and $\Gamma$M. There are two different types of crossings that appear as a result of the folding. The crossing along $\Gamma$M appears between the \textit{p$_x$}- and \textit{p$_y$}-bands, which also cross at the X point. These crossings are a direct result of the symmetry and will always be there, since \textit{p$_x$} and \textit{p$_y$} will always be degenerate at $\Gamma$. The crossing along $\Gamma$X is between the \textit{p$_x$}/\textit{p$_y$}-band and the \textit{p$_z$}-band. This crossing depends on the energetic location of the \textit{p$_z$}-band and can be present or absent in real materials. A Dirac semimetal can be realized if the Fermi level is located within the \textit{p}-manifold. Since the band structure is derived only based on the orbital overlap of identical atoms within a square net, these Dirac crossings are direct results of the bonding within the net.
The multiple crossings usually result in a NLSM.  \textbf{Figure \ref{sqbands}(d)} shows the expected FS for a sole square net, if the Fermi level resides at the \textit{p$_x$}/\textit{p$_y$} band crossing. While so far, there is no example of a material with such a clean square-net derived electronic structure, the diamond-shape of the Fermi surface in \textbf{Figure \ref{sqbands}(d)} resembles the FS observed in ZrSiS and related materials quite well. The difference is that in the real materials, the diamond does not extend all the way to the X point, since the band degeneracy at X is lowered in energy (see \textbf{Figure \ref{strucs}}). But due to additional crossings along $\Gamma$X, the shape of a diamond remains.

\subsection{The square net and non-symmorphic symmetry}

In 2015 \textit{Young} and \textit{Kane} used the square net as an example to predict 2D Dirac semimetals based on non-symmorphic symmetry (NSS) operations \cite{young2015dirac}. These are symmetry operations that contain a translational symmetry element, such as a glide plane or a screw axis. They force bands to be degenerate at BZ boundaries. We can easily rationalize this by considering that the BZ folding in PbFCl compounds is a consequence of NSS. Due to the different packing densities and the stacking of the \textit{M}(\textit{Z}) and \textit{X} layers, there is a M$_z$ glide mirror in the crystal structure, causing the compounds to be non-symmorphic (NS). The NSS elements force the bands to be degenerate at the edges of the BZ, i.e. at the M and X points and along the XM line. This is easy to understand since the degeneracy is enforced at the BZ folds. We can see the degeneracies in the folded band structure shown in \textbf{Figure \ref{sqbands}(c)}. While in the tight-binding (TB) model, the degeneracy at X appears at the same energy than the crossing along the $\Gamma$M line, in real materials these energies are usually not the same (i.e. they lie above and below the Fermi level in ZrSiS, see \textbf{Figure \ref{strucs}(c)}). Unlike the band crossings in graphene and other 2D TSMs, band degeneracies protected by NSS are stable (As well as the crossings along $\Gamma$M) towards SOC. The proposal by \textit{Young} and \textit{Kane} opens a way to create true, ungapped 2D Dirac semimetals in the presence of SOC, which is otherwise not possible. 
PbFCl-type materials partially match the prediction by \textit{Young} and \textit{Kane} and have been investigated in this regard \cite{young2015dirac,schoop2015dirac,topp2016non,topp2017surface,topp2017effect,schoop2017tunable,guan2017two,chen2017dirac,habe2017tunneling}.

\subsection{Tuning the Fermi level}

If the Fermi level is located at any of the crossing points depends on the electron count. For example, the compound PbFCl itself is a trivial insulator \cite{tremel1987square}. The reason is that fluorine, which is occupying the $4^4$-net position, has a filled \textit{p}-shell, since it will have a charge of -1 due to its high electronegativity. Thus, PbFCl is an ionic compound with no in-network F-F bonding. In ZrSiS, Si occupies the $4^4$-net position and its \textit{p}-shell is not filled. If we count electrons in ZrSiS, we would tentatively assign Zr an oxidation state of +4 and S one of -2  \cite{nuss2006geometric}. This would leave us with a -2 oxidation state for Si, which is not enough to fill the \textit{p}-bands. Thus, the Fermi level could be located at the Dirac points. However, \textit{Tremel} and \textit{Hoffmann} already noticed that if the square-net band structure only consists of \textit{s}- and partially filled \textit{p}-bands, as in our model, the square net prefers to distort, which creates a band gap and the material becomes a trivial semiconductor \cite{tremel1987square}. An example is the isoelectronic compound GdPS in which the P-based  $4^4$-net distorts to form zigzag chains, resulting in trivial semiconducting properties.  \textit{Tremel} and \textit{Hoffmann} explained the distortion by arguing that the band crossings in the electronic structure indicate an electronic instability. And indeed, by calculating the crystal orbital overlap population (COOP) curve of GdPS, they found that the Fermi level is located in the anti-bonding region of the P-P overlap curve. They also showed that for a zig-zag chain, this anti-bonding overlap vanishes, while the band crossings disappear.  This suggests that the preferred crystallographic arrangement of P in GdPS is a zig-zag chain. We can rationalize this by considering the Zintl concept \cite{kauzlarich1996chemistry}. Assigning oxidation states to GdPS, results in Gd$^{3+}$P$^-$S$^{2-}$. According to the Zintl concept we would expect P$^-$ to behave as a chalchogen. Elements from column 16 in the periodic table prefer to form two bonds and form zig-zag chains. If we force them into a different bonding arrangement, the normal two electron-two center (2e2c) bonding condition won't be satisfied. Thus there will be fewer electrons to form all four required bonds within the square net, which results in a delocalized bonding network. Therefore, the band structure of a square net with an electron count that cannot accommodate the formation of four bonds, should be related to the band structure of graphene. 
According to Zintl, a distortion should also be expected in ZrSiS. Si$^{2-}$ in ZrSiS is isoelectronic to P$^-$ in GdPS and the Fermi level is located in the anti-bonding region of the Si-Si COOP curve \cite{tremel1987square}. So why does ZrSiS not distort but keeps the square net structure? \textit{Tremel} and \textit{Hoffmann} argued that if \textit{d}-bands are in the vicinity of the dispersive, square-net derived \textit{p}-bands, they can stabilize the square-net structure \cite{tremel1987square}. We can also say that the Zr \textit{d}- and Si \textit{p}-overlap is strongly bonding and overcompensates the anti-bonding Si-Si overlap. And indeed, if the orbital contributions are projected on the individual bands in ZrSiS, it is visible that the steep Dirac bands are mainly composed of Si \textit{p}- and Zr \textit{d}-bands \cite{schoop2015dirac}. We can view the Si network in ZrSiS as an inorganic analog of graphene; in this case the distortion is not prohibited by aromatic stability but by \textit{d}-bands that overlap with the \textit{s}- and \textit{p}-bands. 

We can conclude that the structural motif of a square net facilitates an electronic structure containing graphene-like linearly dispersed bands. In order for the Fermi level to be located at the crossing point, we need an electron count different from the noble gas configuration, leaving the \textit{p}-states partially filled,  allowing the electrons to be delocalized over the square net. \textit{Nuss}, \textit{Wedig} and \textit{Jansen} showed with electron localization function (ELF) calculations, that Si indeed behaves like a chalchogen in ZrSiS, it uses two electrons for bonding and has two lone pairs \cite{nuss2006geometric}. Since it forms four bonds with two electrons, there are only 0.5 electrons available per bond and regular 2e2c bonding is not possible.

This allows us to define a recipe for identifying square-net materials that should be TSMs:  we should search for square-net compounds with an electron count of 6 electrons per $4^4$ net-atom.
Since the square net is inherently unstable with this electron count and prefers to undergo a Peierls distortion if only \textit{s}- and \textit{p}-orbitals play a role, it is important to have neighboring \textit{d}-bands available to stabilize the square net. In the next section, we will discuss the process of identifying the best candidate materials in more detail.

\section{STRUCTURE-PROPERTY RELATIONS}

\begin{figure}[h]
\includegraphics[width=4in]{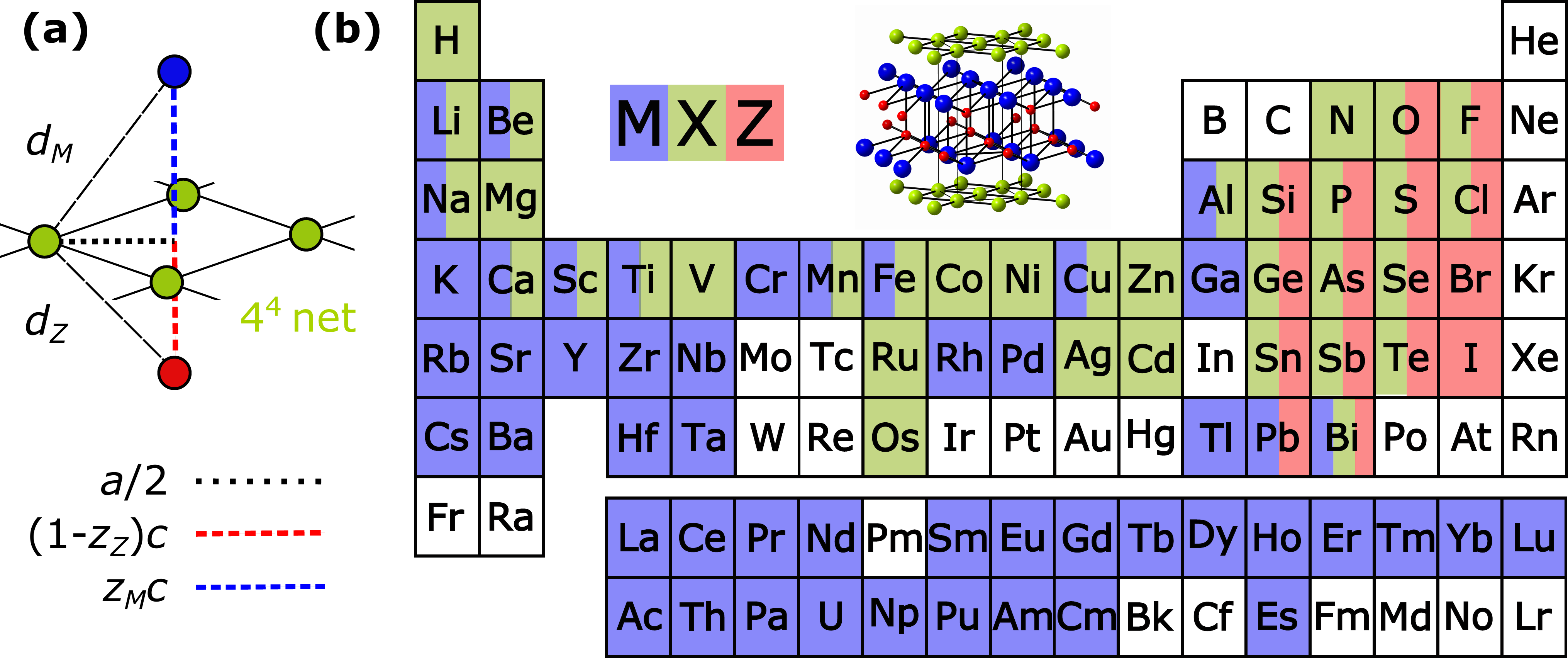}
\caption{(a) The schematic shows the 4\textsuperscript{4} net and its neighboring \textit{Z} and \textit{M} atoms. The determination of the parameters \textit{d\textsubscript{M}} and \textit{d\textsubscript{Z}} relies on \textit{a}, \textit{c} and \textit{$z_M$} or \textit{$z_Z$} respectively. (b) Color coded periodic table for \textit{MXZ} compounds crystallizing in SG \textit{P}4/\textit{nmm} with \textit{M} (blue), \textit{X} (green) and \textit{Z} (red).}
\label{PSE}
\end{figure}

The subtle crystallographic details of square-net materials have a significant influence on their electronic structure. We will therefore devote this section to the crystallogrpahic aspects of square-net materials. We will focus on the PbFCl family of compounds, which can be sub-divided into individual structure types (PbFCl, PrOI, ZrSiS, Cu\textsubscript{2}Sb). They all crystallize in the space group \textit{P}4/\textit{nmm} and the atoms occupy Wyckoff sequence \textit{c}2\textit{a}, stacked along [001]. The subtle differences lie in stacking order, stacking distance and the degree of buckling within the rock salt-like sub-structure. While we account for the differences, we will group all these structures into one family for simplicity. For a detailed overview on these structure types and the compounds adopting them, we refer to crystallography focused reviews \cite{nuss2006geometric,charkin2007crystallographic}.
For the purpose of this discussion, we will describe the PbFCl structure type as a layered structure with two distinguishable layers per unit cell. The layers consist of three chemically different atoms \textit{M}, \textit{X} and \textit{Z}. The \textit{M} and \textit{Z} atoms occupy the 2\textit{c} Wyckoff sites (1/4, 1/4, \textit{z\textsubscript{M}}, 3/4, 3/4, -\textit{z\textsubscript{M})} and (1/4, 1/4, \textit{z\textsubscript{Z}}, 3/4, 3/4, -\textit{z\textsubscript{Z}}) respectively. Thus these two atoms each contain one variable coordinate, and its exact value is very important for understanding the subtle differences that appear within this family of compounds. The resulting double layer is reminiscent of a rock salt-like layer if \textit{z\textsubscript{M}} = (1-\textit{z\textsubscript{Z}}). In reality, the values for \textit{z} differ and the rock salt-like layer buckles \cite{nuss2006geometric}. 
The buckled layer can be seen as four staggered stacked square nets of either \textit{M} or \textit{Z} with a mesh size of \textit{a}.
The atomic order along the \textit{c}-axis can be either \textit{MZZM} or \textit{ZMMZ} depending on the electronegativity and electron configurations of the \textit{X} atoms. The \textit{X} atoms form the 4\textsuperscript{4} square net (Wyckoff position 2\textit{a}). This net is twisted by 45$^\circ$ in respect to the square nets formed by \textit{M} and \textit{Z}. The in-plane distance between the square \textit{X} atoms (\textit{d}\textsubscript{sq})  is  \( \frac{\sqrt[]{2}a}{2} \) and therefore, the \textit{X} layer is twice as dense as the \textit{M} or \textit{Z} layers.
To compare the structures and to find structure-to-property relationships, we define the structural parameter \textit{d}. This parameter is the distance between the 4\textsuperscript{4} net and its closest neighbor. It can either be \textit{d\textsubscript{M}} or \textit{d\textsubscript{Z}} depending on the values of \textit{z\textsubscript{M}} and \textit{z\textsubscript{Z}}. \textbf{Figure \ref{PSE} (a)} shows a graphical explanation for the determination of \textit{d}; it can be easily calculated with the lattice constants \textit{a}, \textit{c} and the variable coordinate of the 2\textit{c} Wyckoff position \textit{z}. We assume the 4\textsuperscript{4} net to be isolated from the rock salt-like layer, if \textit{d} is larger than the interatomic distance within the 4\textsuperscript{4} net \textit{d}\textsubscript{sq} and argue these compounds to have more 2D than 3D character. The electronic structure of a compound with $d_{\mathrm{sq}}/d < 1$ should thus potentially contain the features of the 2D square net. 

In order to understand how structural details affect the electronic structure, and when Dirac semimetals can appear, we exploited the structure data from the International Crystallographic Database (ICSD) \cite{ICSD} to compile a structural map of the known PbFCl-type materials. \textbf{Figure \ref{PSE} (b)} shows which elements are known to occupy the different  \textit{M} (blue), \textit{X} (green) and \textit{Z} (red) sites in PbFCl-type materials. The over 500  known compounds show a variety of electronic properties depending on their bonding character (ionic, metallic, covalent) and their electron count \cite{nuss2006geometric}. The \textit{M} position is usually occupied by the least electronegative element and works as an electron donor. Accordingly, the \textit{Z} position is occupied by an electron acceptor and the role of the atom at the \textit{X} position is ambivalent.Taking this ambivalence in account, we can formulate two different valence electron distributions in \textit{MXZ} phases: \textit{M\textsuperscript{x+}X\textsuperscript{y+}Z\textsuperscript{(x+y)-}} if X is an electron donor, or \textit{M\textsuperscript{x+}X\textsuperscript{y-}Z\textsuperscript{(x-y)-}} if it is an acceptor.

\begin{figure}[h]
\includegraphics[width=5in]{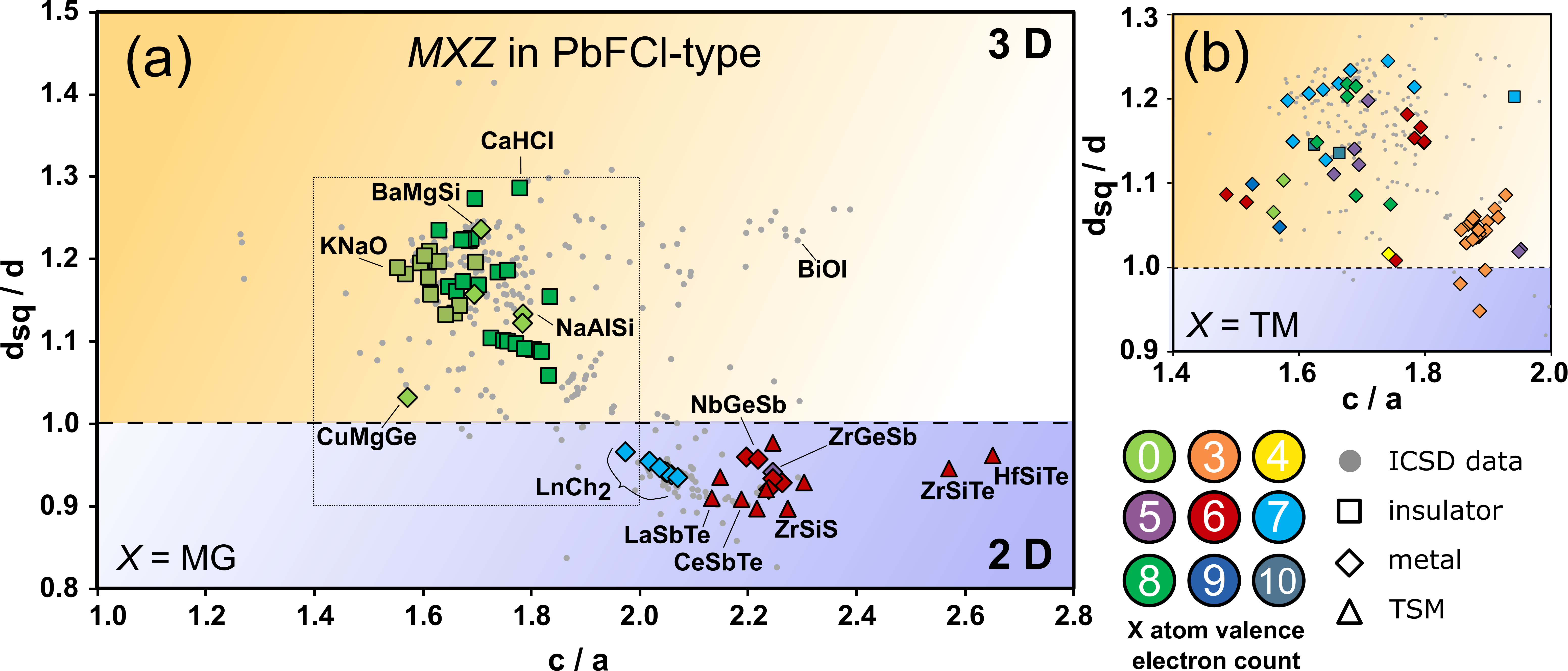}
\caption{Plot of \textit{d}\textsubscript{sq}/\textit{d} vs \textit{c}/\textit{a} for different PbFCl-type compounds with \textit{X} being a MG element (a) or a TM (b). The shapes represent insulating (square), metallic (diamond) or clean TSM (triangle) phases. Note that phases that contain linear band crossings convoluted with other states are labeled as metallic here. Color coding represents the electron count of the $4^4$ net atoms.} 
\label{DATA}
\end{figure}

\textbf{Figure \ref{DATA}} shows plots of the distance ratio \textit{d}\textsubscript{sq}/\textit{d} versus \textit{c}/\textit{a}. The 2D character of the compounds increases towards the lower right side of the plot. We created two of these structure maps. In \textbf{Figure \ref{DATA}} (a) only compounds with a main group (MG) element on the \textit{X}  position are shown, while (b) shows only compounds where \textit{X} is a transition metal (TM). The gray data points show all reasonable compounds listed in the ICSD database as reference, even if the electronic structure has not been reported.  Meanwhile the colors and shapes of the data points indicate the electron count of the \textit{X} atoms and the electronic properties of the compound, respectively. We used the \textit{materialsproject.org} database \cite{Jain2013} to extract information about the electronic structure of the compounds, as long as the structural parameters \textit{a} and \textit{c} did not deviate more than 0.1$\,\mathrm{\AA}$ from the values reported in the ICSD.

The plot results in a structure and property map of PbFCl-type compounds, where different regions in the plot can be distinguished and connected to structural and electronic differences. Noteworthy is that compounds with similar electron counts appear to cluster in the same region of the map. 
We will start by discussing the compounds in which \textit{X} is a MG element (\textbf{Figure \ref{DATA} (a)}). The main body of the data points is located in the upper left area of the plot, which is the region where the crystal structure is more 3D. These compounds are mostly ionic insulators and charge-balanced, the formal electron count is eight or zero valence electrons per \textit{X} atom. The \textit{X}-position is either occupied by halide or oxide ions (as in PbFCl or BiOCl), or by alkali or alkaline earth cations mainly Li\textsuperscript{+} and Na\textsuperscript{+} (as in NaLiS and RbNaO). An exception are metallic compounds where \textit{X} is occupied by Mg\textsuperscript{2+} or Al\textsuperscript{3+}. Examples are the superconductor NaAlSi, as well as NaAlGe and the series of BaMgCh (Ch = chalcogenide) compounds. The metallic character of these charge-balanced compounds arises due to overlap between the filled \textit{p}-shell with the next empty shell. Note that such compounds can still be ``atomic" Dirac semimetals regardless the nature of the bonding character \cite{KMgBi-paper}. 

There is one region in the plot that clearly separates from the rest: The region where \textit{d}\textsubscript{sq}/\textit{d} is smaller than 1, compounds that appear there have a higher 2D character. All compounds in that region are not electron precise, they have an electron deficiency on the \textit{X} position. To compensate the missing electrons, the atoms form bonds, resulting in shorter distances within the 4\textsuperscript{4} net and a decreased distance ratio. The electronic structures of these compounds should exhibit the features of a 2D square net. They can be divided into two different groups based on the electron count of \textit{X}. Compounds such as LnCh\textsubscript{2} (Ln = lanthanides, Ch = chalcogenides) with formally 7 electrons per square-net atom appear on the left side of the plot, while compounds such as ZrSiS with 6 electrons per square-net atom appear on the right side. Note that all confirmed TSMs appear in the same region than ZrSiS. One exception is ZrGeSb, which formally has 5 valence electrons per square net atom and is located in the 6 electron area of the plot. ZrGeSb is the only example with this electron count and its electronic structure contains linear bands convoluted with conventional parabolic ones. In addition to the 7 valance electron  compounds, actinoide (An) compounds such as AnCh\textsubscript{2}, AnPnCh and AnPn\textsubscript{2} appear in that part of the structural map, but are not part of the analysis.

The band structures of the electron deficient square-net species all show crossings of steep bands at the Fermi level, but as in ZrGeSb, other features can spoil them. In that case they are labeled as metallic in the structural map. 
 Compounds with 7 valence electrons per X atom have a rather metallic and convoluted electronic structure, but if one chalcogen is substituted by a pnictide (Pn), relatively clean Dirac crossings appear at the Fermi level. These compounds have 6 valence electrons on the \textit{X} site, following the formula LnPnCh, and an increased \textit{c}/\textit{a} ratio. 

All compounds with TSM band structures are isoelectronic to ZrSiS. 
Besides the formal electron count, the charge of the \textit{M} atom  influences the electronic properties as well. Despite formally having 6 electrons per silicon, NbSiAs is not a clean TSM. Its electronic structure contains Dirac crossings at the Fermi level, but additional bands appear. We can rationalize this by trends in formal oxidation states. While 3+ cations are often really charged 3+, higher oxidation states are more a formality than reality. There are not many known PbFCl-type compounds with 3+ cation (a few exceptions are CeSbTe and LaSbTe and some very metallic or charge-balanced ionic compounds). As we discussed in the previous section, with the example of GdPS, compounds with trivalent \textit{M} cations often distort. Compounds with a tetravalent \textit{M} cation, such as ZrSiS are stabilized by the \textit{M} \textit{d}-states. Thus the \textit{M} cation is not really charged 4+, but it is close enough to that oxidation state that not too many \textit{d}-bands interfere with the square-net derived bands at the Fermi level. If \textit{M} has the formal charge of +5 however, this effect becomes too large, and \textit{M} \textit{d}-bands interfere with the linear bands at the Fermi level.

\textbf{Figure \ref{DATA} (b)} shows the structural data of compounds where \textit{X} is a TM. All these compounds are metallic, with many (often not very dispersive) bands crossing the Fermi level. The main body of the data is located on the top half of the map, where the crystal structure is more 3D. The only seperate cluster of compounds with a more 2D character contain early TMs (Sc, Ti, V) as \textit{X} atoms with a low electron count.  
 While the metallic band structures often feature band crossings, no clean Dirac semimetal where \textit{X} is a TM was found. Since the square-net band structure was modeled for \textit{p} rather than \textit{d} electrons, this is not very surprising.

We can derive several important messages from the structural investigation of the PbFCl-type compounds: All \textit{MXZ} compounds that feature an electronic structure reminiscent of a clean TSM exhibit a large \textit{c}/\textit{a} ratio as well as isolated $4^4$ square nets (i.e. the distance ratio is smaller than 1). The atoms that form these nets have only partially filled \textit{p}-shells and are isoelectronic to chalcogene atoms. Fewer or extra electrons in the square-net lead to interfering band crossings at the Fermi level and cause the compounds to be rather metallic. Comparing compounds with different oxidation states on the \textit{M} atoms such as Gd\textsuperscript{3+}, Zr\textsuperscript{4+} or Nb\textsuperscript{5+} leads to the conclusion that an oxidation state of 4+ is optimal for \textit{d}-band stabilization of the Dirac crossings without polluting the Fermi level.

The results of our analysis for \textit{MXZ} compounds in the PbFCl-structure type family should be transferable to other structures containing 4\textsuperscript{4} square nets. To give a detailed analysis of all compounds is out of the scope of this review. A quick analysis nevertheless revealed similar trends for \textit{MXZ\textsubscript{2}} compounds. 

\begin{figure}[h]
\includegraphics[width=3.5in]{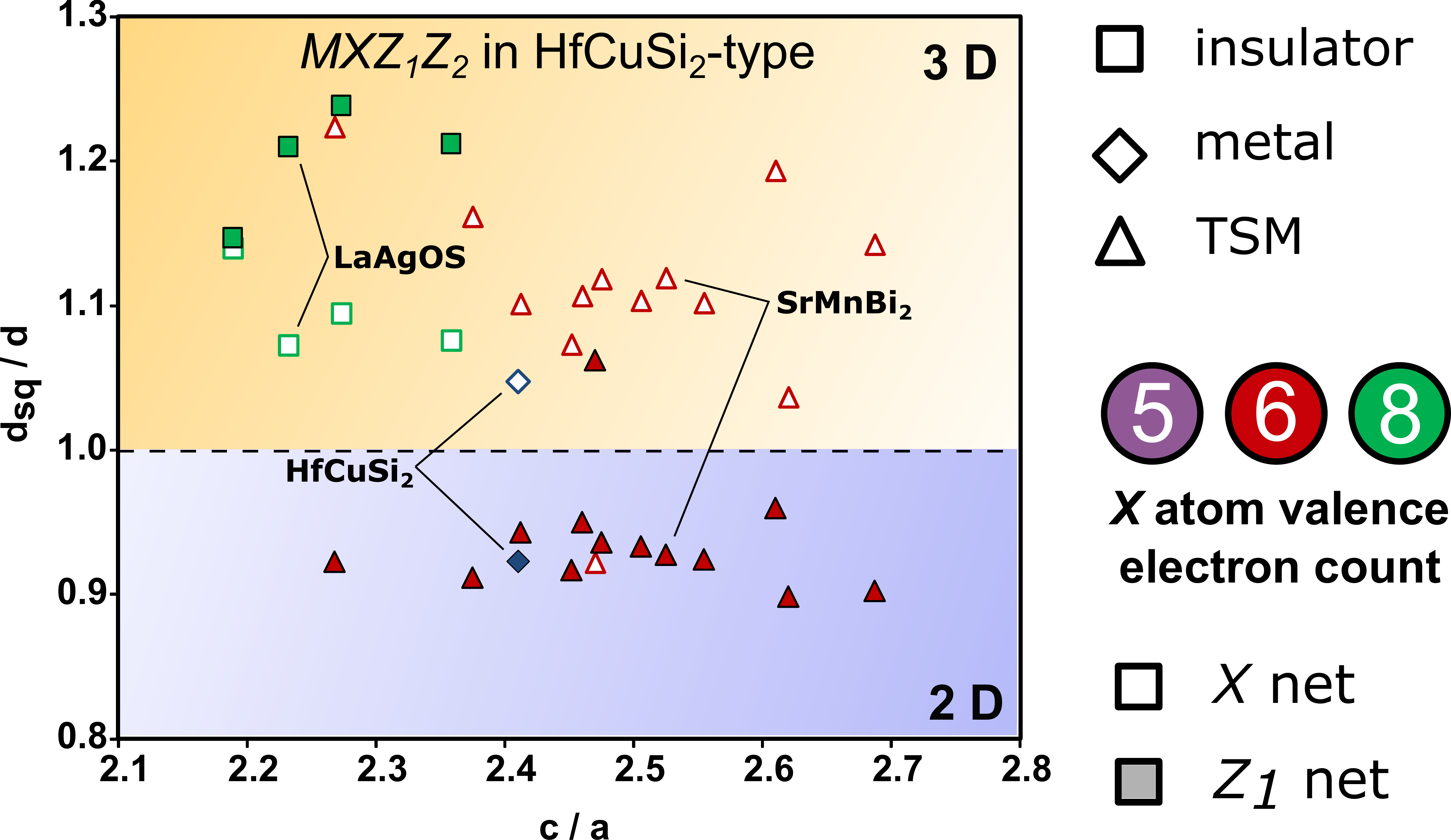}
\caption{Plot of \textit{d}\textsubscript{sq}/\textit{d} vs \textit{c}/\textit{a} for different \textit{MX$Z_1Z_2$} compounds. The shapes represent insulating (square), metallic (diamond) or TSM (triangle) phases. These compounds exhibit two $4^4$ nets of \textit{X} (empty shapes) or \textit{$Z_1$} (filled shapes) atoms respectively. Color coding represents the electron count of the \textit{$Z_1$} $4^4$ net atoms.}
\label{MXZ2chart}
\end{figure}

In the fashion of the structural map for \textit{MXZ} compounds, \textbf{Figure \ref{MXZ2chart}} shows data points for reported \textit{MXZ\textsubscript{2}}-type TSMs and some insulating, charge-balanced representatives with very similar structure type (\textit{MX$Z_1Z_2$} = LaAgOS, BiCuOSe, BiCuOTe, SrCuFS in the ZrCuSiAs structure type in \textit{P4/nmm}). The parameters where adjusted and generalized for the different structure types and crystal systems. All these compounds, pseudo-ternary or quarternary, exhibit a second $4^4$ net. \textbf{Figure \ref{MXZ2chart}} shows the \textit{d}-values for the \textit{X} and \textit{$Z_1$} net.
Similar as in \textit{MXZ} phases, the compounds that are TSMs contain a square net composed of a MG element with 6 valance electrons per atom and appear in the 2D area of the plot. Meanwhile, the \textit{X} square net consists of TM atoms and has a more 3D character. For the insulating, charge-balanced compounds both $4^4$ nets appear in the 3D area of the structural map.

While  maps were constructed for PbFCl-type phases structural before \cite{nuss2006geometric,charkin2007crystallographic}, they have not yet been connected to electronic properties. The structural map we derived for \textit{MXZ} compounds revealed a relation between structure parameters and topological properties. 
We can conclude that an estimate of the electronic properties of square-net materials can be made solely based on crystallographic parameters. In other words we defined a ``tolerance factor" $t = \frac{d_{sq}}{d} = \frac{\sqrt{2} a}{2 \sqrt{(zc)^2 + a^2/4}} < 1$ for topology in PbFCl-type  materials.

\section{PROPERTIES OF TOPOLOGICAL SQUARE NET MATERIALS}
\subsection{\textit{MXZ} phases}

\begin{figure}[h]
\includegraphics[width=5in]{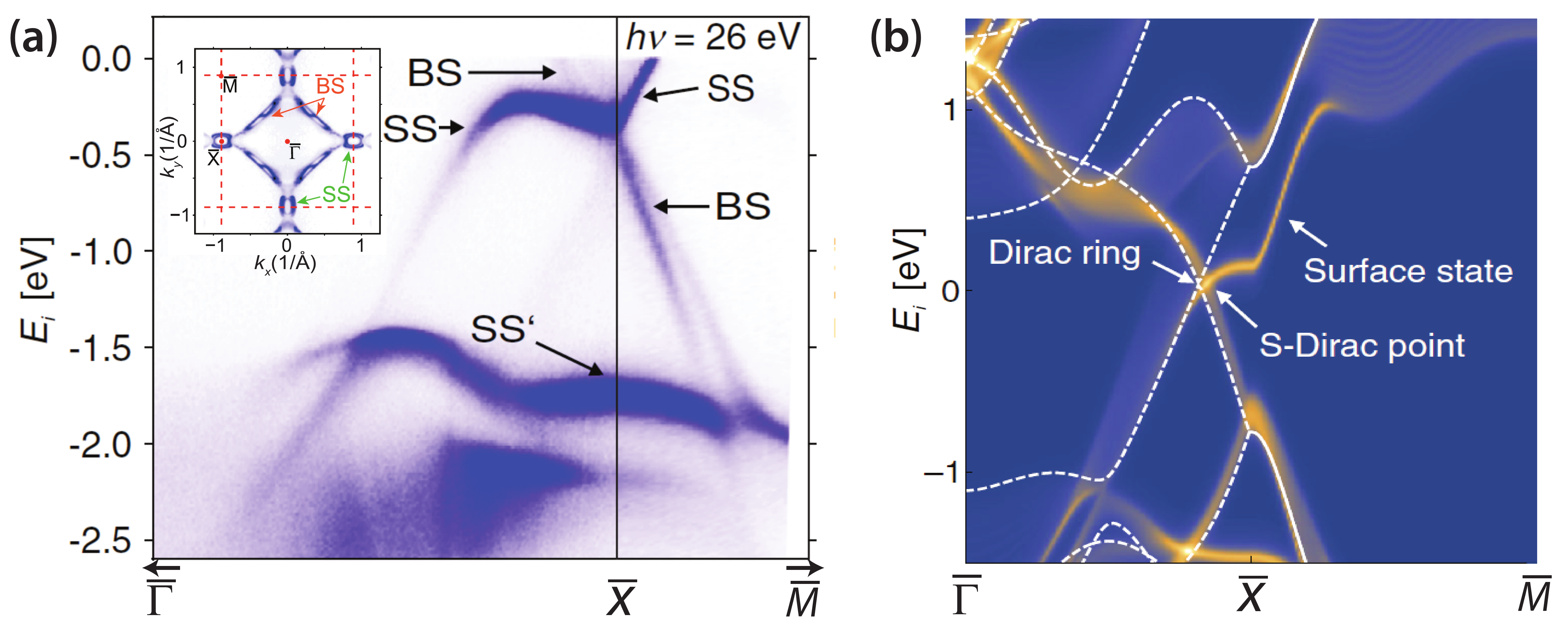}
\caption{ (a) ARPES resolved surface states (SS) on ZrSiS (001) surface. BS stands for bulk states. Inset shows the measured diamond-shaped FS. (b) Surface band structure based on a 26-band TB model. The symmetry reduction at the surface creates a two-dimensional surface floating band. Figures are reprinted from \cite{topp2017surface} and \cite{chen2017dirac}}.
\label{ZrSiS_surface}
\end{figure}

As shown by our structural map, ZrSiS and its closely related compounds with \textit{M} = (Zr, Hf), \textit{X} = (Si, Ge, Sn) and \textit{Z} = (S, Se, Te), stand out rather uniquely as very clean Dirac semimetals within the PbFCl  family. Thus unsurprisingly, ZrSiS is the most heavily studied compound in this class. Its electronic structure was investigated multiple times with ARPES \cite{schoop2015dirac,neupane2016observation,wang2016evidence, chen2017dirac,topp2017effect, topp2017surface, sankar2017crystal,fu2017observation}, STM \cite{lodge2017observation,butler2017quasiparticle} or optically \cite{schilling2017flat}. These studies established ZrSiS as a novel Dirac NLSM and confirmed the large energy range of linear bands, without interference from other bulk bands.  ARPES also revealed the existence of surface states (\textbf{Figure \ref{ZrSiS_surface}}). The origin of these surface states was later attributed to the symmetry reduction at the surface of NS compounds \cite{topp2017surface}. At the surface, the translation symmetry along the surface normal direction is broken. Therefore, the original NS glide-mirror symmetry in space group \textit{P4/nmm} is lifted, and it is reduced to the symmorphic wall paper group of \textit{P4mm}. This causes the original symmetry-enforced degenerate bulk bands at $\overline{\mathrm{X}}$ to lift their degeneracy and creates \textit{surface floating band}. This explanation of the surface bands only involves symmetry arguments, therefore the existence of similar surface floating bands is expected in other layered NS compounds. This is indeed the case in ZrSiTe \cite{topp2016non} and HfSiS \cite{takane2016dirac}.

STM studies elucidated the electronic structure by identifying the scattering vectors that give rise to the observed constant-energy quasi-particle interference (QPI) patterns. \textit{Lodge} et al. \cite{lodge2017observation} were able to determine the energy position of the line nodes along $\Gamma−M$, and the Fermi velocity in the conduction band. 
Independently, \textit{Butler} et al. \cite{butler2017quasiparticle} investigated the QPI dependence on the both Zr and S defects on ZrSiS. They found that the quasiparticle scattering possesses unusual selectivity depending on the specific type of point-defect scattering center.

Besides direct probes of the electronic structures, the properties of ZrSiS have also been studied widely by magnetotransport \cite{ali2016butterfly,lv2016extremely,wang2016evidence,matusiak2017thermoelectric,hu2017nearly,singha2017large,pezzini2018unconventional,zhang2018transport, sankar2017crystal}. All these studies reveal multiple FS pockets. Particularly, the Hall resistivity suggests the existence of both electron- and hole-type carriers \cite{singha2017large,zhang2018transport,sankar2017crystal}. A high electron and hole mobility, and an extremely large and unsaturated magnetoresistance, reaching ~1.8 $\times$ $10^5$\% at 9 T and 2 K at an 45$^\circ$ angle between the applied current ($I \parallel a$) and the applied field (90$^\circ$ is $H \parallel c$) was found by several groups \cite{ hu2017nearly,ali2016butterfly,lv2016extremely,singha2017large}. 
Quantum oscillation measurements \cite{wang2016evidence,singha2017large,hu2017nearly} indicate a nontrivial Berry phase.  
One particular interesting feature found in transport studies is the appearance of magnetic breakdown across gaps along the nodal loop \cite{pezzini2018unconventional}. The effective masses associated with these new frequencies are significantly enhanced beyond the conventional band-structure estimates.  This discovery indicates that ZrSiS is likely to be close to a quantum phase transition. Therefore, ZrSiS may provide an ideal material platform on which to explore novel correlated physics associated with the topological nodal loop. 




\begin{table}[hbt]
\caption{\textit{MXZ}-type topological materials and their experimental verification. The following acronyms are used for methods: MT = Magnetotransport, QO = Quantum oscillations, STM = Scanning tunneling microscopy, OC = optical conductivity, UPP = Ultrafast optical pump-probe, RS = Raman spectroscopy. The numbers in the cell represent the references for the work.}
\label{tab1e1}
\centering
\begin{tabular}{c|c|c|c|c|c}
\hline
\cellcolor{yellow!20}Material & \cellcolor{yellow!20}ARPES &  \cellcolor{yellow!20}MT and QO & \cellcolor{yellow!20} STM & \cellcolor{yellow!20} OC and UPP & \cellcolor{yellow!20} RS \\
\hline

  ZrSiS & \makecell{\cite{schoop2015dirac,neupane2016observation,wang2016evidence,chen2017dirac}\\ \cite{topp2017effect, topp2017surface, sankar2017crystal,fu2017observation}} &  \makecell{
   \cite{ali2016butterfly,lv2016extremely,wang2016evidence,matusiak2017thermoelectric,hu2017nearly}\\ \cite{singha2017large,pezzini2018unconventional,zhang2018transport, sankar2017crystal}} &    \cite{lodge2017observation,butler2017quasiparticle} &\cite{schilling2017flat,Weber2017similar}    
   & \cite{zhou2017lattice,singha2018probing}\\
\hline
ZrSiSe &  & \cite{hu2016evidence,pan2018three}   & \cite{bu2018visualization} & & 
\\
\hline
ZrSiTe & \cite{topp2016non} & \cite{hu2016evidence} & & & \\
\hline
HfSiS & \cite{takane2016dirac,chen2017dirac} & \cite{kumar2017unusual,van2018electron} & & &\\
\hline
ZrGeS &  &  \cite{hu2017quantum} & & & \\
\hline
ZrGeSe &  & \cite{hu2017quantum} & & &\\
\hline
ZrGeTe & \cite{hosen2018observation} & \cite{hu2017quantum} & & &\\
\hline
ZrSnTe &  \cite{lou2016emergence} & \cite{hu2018quantum} & & & \\
\hline
CeSbTe & \cite{schoop2017tunable,topp2017effect} &   & & &  \\
\hline
\end{tabular}
\label{tabZrSiS}
\end{table}

Besides the immensely studied ZrSiS, its sister compounds have also been actively investigated for their Dirac nodal-line electronic structure. Table \ref{tabZrSiS} gives an overview of all ZrSiS-type materials that have been experimentally investigated. While many materials exhibit a very similar electronic structure to ZrSiS, there are some differences. For example, in the case of HfSiS, ARPES revealed unique \textit{Dirac-node arc} surface states \cite{takane2016dirac}. The origin is so far unclear and its consequence to the overall band topology is still to be explored.
In addition, magnetotransport and quantum oscillation studies have been performed on HfSiS \cite{kumar2017unusual,van2018electron}, ZrSiSe \cite{hu2016evidence,pan2018three} and ZrSiTe \cite{hu2016evidence}. Similarly as in ZrSiS, small effective masses as well as a nontrivial Berry phase were detected. In the case of HfSiS and ZrSiSe, quantum oscillations indicated that the FS enclosing the nodal-line is of 3D character, similar to that of ZrSiS. In contrast, ZrSiTe hardly shows any quantum oscillations for \textit{B}$\parallel$\textit{ab}, suggesting a rather 2D-like Fermi surface. The difference is consistent with their huge difference in the interlayer binding energy, which was calculated to be one order of magnitude smaller in ZrSiTe than in HfSiS and ZrSiS \cite{xu2015two}.
This is closely related to the \textit{c/a}-ratio, which is increased by ~13\,\% in ZrSiTe in comparison to the other ZrSiS-like materials (see also \textbf{Figure \ref{DATA}}).
This effectively creates chemical strain and can be utilized for functionality tuning. For example, \textit{Topp} et al. \cite{topp2016non} realized that the \textit{c/a}-ratio is closely related to the energy location of the NS Dirac cones in these compounds, and ZrSiTe has the ideal \textit{c/a} ratio to shift a NSS protected Dirac crossing to the Fermi level, which was also confirmed by ARPES studies. ZrSiTe is thus considered as the first real material exhibiting an NS Dirac crossing at the Fermi level.

HfSiS was found to possess unique transport properties. It was discovered that electrons can tunnel between electron and hole pockets in high-field Shubnikov–de Haas oscillations \cite{van2018electron}. This is unexpected because the enhanced SOC in HfSiS should give rise to a  SOC-induced gap that is 2-3 times larger than that in ZrSiS. The study also represents the first experimental confirmation of the momentum-space Klein tunneling.

CeSbTe stands out uniquely in the family of ZrSiS-type materials, as it is the only magnetic member. It was shown to exhibit a complex magnetic phase diagram \cite{schoop2017tunable}. In its antiferromagnetic phase below \textit{T}$_N$=2.7\,K, the Ce$^{3+}$ moments were found to order ferromagnetically within the \textit{ab} plane, while stacking along the \textit{c}-axis in the sequence of up-down-down-up. In this way, the magnetic unit cell is doubled along \textit{c}-axis. Tunable Weyl and Dirac states are expected based on the fact that there are field-induced magnetic phase transitions for both \textit{B}$\parallel$\textit{c} and \textit{B}$\perp$\textit{c}, which leads to the formation of various fully polarized phases. Since these different magnetic phases are described by different magnetic space groups, a group theoretical analysis suggests that they will have different electronic structures. 
Particularly, in the antiferromagnetic state an eight-fold degenerate point can be found, which has been associated with the appearance of new fermions \cite{schoop2017tunable,bradlyn2016beyond}. Overall, CeSbTe is predicted to host many different topological features in its electronic structure. As a single material, it allows access to many different magnetic groups, thus providing a versatile platform to study the effect of magnetic-order-induced symmetry change on different types of Dirac crossings, as well as on higher-order degeneracies that result from NSS.

\subsection{\textit{MXZ$_2$} phases}

There are three different square-net-based crystal structures with the \textit{MXZ$_2$} formula. Two of these structures are represented by the CaSmP$_2$ (space group \textit{I4/mmm}) and the HfCuSi$_2$ (space group \textit{P4/nmm}) structure types and have been introduced in section 2. In addition, space group \textit{Pnma} can also host square-net-based \textit{MXZ$_2$} phases (SrMnSb$_2$ structure type), although in this case the square-net is slightly distorted and resulting in an orthorhombic symmetry. As listed in \textbf {Table \ref{tab1e2}},  materials with nontrivial band topology have been reported in all these three crystal structures. 

\begin{table}[hbt]
\caption{\textit{MXZ}-type topological materials and their experimental verification. The following acronyms are used for methods: MT = Magnetotransport, QO = Quantum oscillations, ND = Neutron Diffraction, NIS = Neutron Inelastic Scattering, OC = optical conductivity, UPP = Ultrafast optical pump-probe, RS = Raman spectroscopy. The numbers in the cell represent the references for the work.}
\label{tab1e1}
\begin{center}
\begin{tabular}{c|c|c|c|c|c}
\hline
\cellcolor{yellow!20}Material & \cellcolor{yellow!20}ARPES &  \cellcolor{yellow!20}MT and QO & \cellcolor{yellow!20} ND and NIS & \cellcolor{yellow!20} OC and UPP & \cellcolor{yellow!20} RS \\
\hline

SrMnBi$_2$ & \cite{park2011anisotropic,jia2014observation,feng2014strong} & \cite{park2011anisotropic,wang2011quantum,wang2012large,he2012giant, jo2014valley} & \cite{guo2014coupling,rahn2017spin} & \cite{park2017electrodynamic,ishida2016revealing} & \cite{zhang2016interplay} \\
\hline
CaMnBi$_2$ & \cite{feng2014strong} & \cite{wang2012large,he2012giant,wang2012two, wang2016interlayer} & \cite{guo2014coupling,rahn2017spin} & & \cite{zhang2016interplay} \\
\hline
EuMnBi$_2$ & \cite{borisenko2015time} &\cite{may2014effect,masuda2016quantum} & & &   \\
\hline
YbMnBi$_2^{\dagger}$ & \cite{borisenko2015time}& \cite{wang2016magnetotransport,liu2017unusual} & & \cite{chinotti2016electrodynamic,chaudhuri2017optical}& \\
\hline
YbMnSb$_2$ & \cite{kealhofer2018observation} & \cite{kealhofer2018observation,wang2018quantum}  & & &   \\
\hline
BaMnBi$_2$ & & \cite{li2016electron,wang2016large} & & &   \\
\hline
BaMnSb$_2$ & &  \cite{liu2016nearly,huang2017nontrivial} & & &   \\
\hline
LaAgSb$_2$ & \cite{shi2016observation,arakane2007electronic} & \makecell{\cite{myers1999systematic,myers1999haas,wang2012multiband} \\ \cite{mun2011thermoelectric,bud2008thermal}} & & \cite{chen2017revealing} &\cite{chen2017revealing} \\
\hline
LaAgBi$_2$ & & \cite{wang2013quasi} & & & \\
\hline
BaZnBi$_2^{\dagger}$ & & \cite{zhao2018quantum,ren2018absence,wang2017magneto} & & &  \\
\hline
\makecell{Sr$_{1-y}$Mn$_{1-z}$Sb$_2$} & &  \cite{liu2017magnetic}& & \cite{Weber2017similar} & \\
\hline
CaMnSb$_2$ &   & \cite{he2017quasi} & & &   \\
\hline
SrZnSb$_2$ & &  \cite{wang2012large_SrZnSb2,liu2018quantum} & & &  \\
\hline
\end{tabular}
\end{center}
\begin{tabnote}
$^{\dagger}$Currently still under debate
\end{tabnote}
\label{tab1e2}
\end{table}

The crystal structures of SrMnBi$_2$ and CaMnBi$_2$ were introduced before and can be found in \textbf{Figure \ref{strucs} (a) and (b)}. 
 Besides the structural differences to the ZrSiS family, SrMnBi$_2$ and CaMnBi$_2$ also differ in their magnetic properties. The Mn atoms are in a +2 oxidation state, as shown by DFT calculations \cite{park2011anisotropic,lee2013anisotropic,guo2014coupling} and neutron diffraction measurements \cite{guo2014coupling}. Due to strong Hund's coupling,  Mn$^{2+}$ is in a high-spin state. We note that the electron counting in SrMnBi$_2$ and CaMnBi$_2$ is less straightforward compared to materials in PbFCl phase.  Bi is in a different oxidation state on the two different sites, the puckered rock salt layer and the $4^4$-net. It is charged -3 in the former case and -1 in the latter case. Therefore, the Bi in the $4^4$-net has effectively 6 electrons and is isoelectronic to Si in ZrSiS. In SrMnBi$_2$ (CaMnBi$_2$), the Mn sublattice orders antiferromagnetically below $\sim$295\,K ($\sim$305\,K) with the magnetic moments aligned parallel to the \textit{c}-axis \cite{park2011anisotropic,guo2014coupling, rahn2017spin}. Because of the large spin polarization of the Mn 3\textit{d}-electrons, the Mn 3\textit{d}-bands are located away from the Fermi level. Therefore, the electronic states near the Fermi level (\textit{E}$_\mathrm{F}$) have mainly Bi \textit{p$_x$} and \textit{p$_y$} character. In this sense, the Dirac fermions residing in SrMnBi$_2$ are directly related to the Bi square net, rather than the Mn square net. 

\begin{figure}[h]
\includegraphics[width=5in]{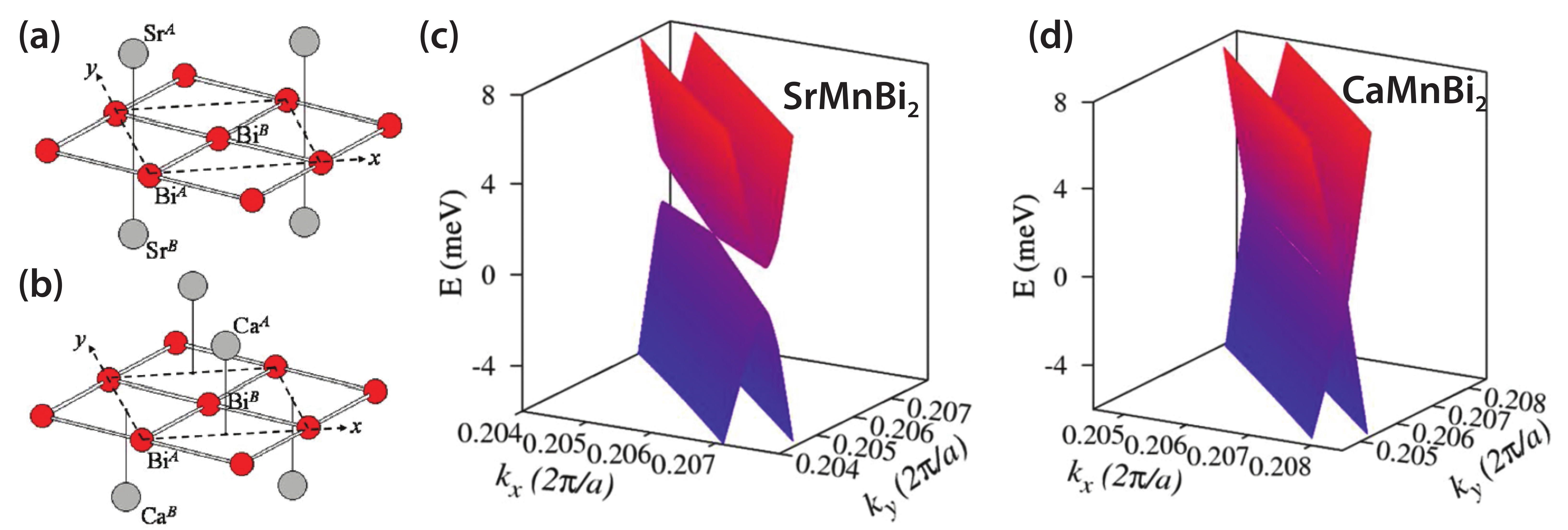}
\caption{(a) and (b) Different arrangements of the neighboring Sr or Ca atoms in respect to the Bi square net in SrMnBi$_2$ and CaMnBi$_2$, respectively. (c) and (d) Energy surface plot \textit{E}($\textit{k}_x$,)$\textit{k}_y$ near the Dirac point from TB band structure calculations for SrMnBi$_2$ and  CaMnBi$_2$. Note: the TB model considered all \textit{d} orbitals of Sr/Ca in hybridization with Bi \textit{p$_x$} and \textit{p$_y$} orbitals. Figures are reprinted from \cite{lee2013anisotropic}.}
\label{SrMnBi2_CaMnBi2}
\end{figure}

One distinguished feature of SrMnBi$_2$ is that it is found to host highly anisotropic Dirac fermions, with a large difference (a factor of ~8 between the maximum and minimum \cite{park2011anisotropic}) in the momentum-dependent Fermi velocity (v$_\mathrm{F}$), in contrast to the generally isotropic Dirac cone in graphene and TIs. The reason for such an anisotropy is closely related to the chemical environment of the Bi square net. As shown in \textbf{Figure \ref{strucs} (a) and (b)}, the Bi square nets are sandwiched between two Sr layers of half packing density. When projected along the \textit{c}-axis, the two adjacent Sr planes coincide (\textbf{Figure \ref{SrMnBi2_CaMnBi2}} (a)). This is different from CaMnBi$_2$, where the Ca atoms are staggered along the \textit{c}-axis (\textbf{Figure \ref{SrMnBi2_CaMnBi2}} (b)). It is this different layering configuration of alkaline earth atoms that results in the body-centered tetragonal Bravais lattice for SrMnBi$_2$ and the primitive one for CaMnBi$_2$, and eventually causes different band structures for these two TSMs (see \textbf{Figure \ref{SrMnBi2_CaMnBi2} (c) and (d)} for the electronic structures).

The electronic structures of SrMnBi$_2$ and CaMnBi$_2$ have been investigated by both first principle calculations \cite{park2011anisotropic,wang2011layered,lee2013anisotropic, guo2014coupling, wang2012large} and TB models \cite{lee2013anisotropic}. In the case of SrMnBi$_2$, the degeneracy along the Dirac nodal-line is lifted at all momenta except a single point, while in CaMnBi$_2$ it is retained. 
To understand this difference let's start with the electronic structure of an isolated Bi square net. The ${p_{x,y}}$-bands of the Bi square net are expected to form the typical nodal-line FS. 
However, this does not explain the appearance of a single anisotropic Dirac cone in SrMnBi$_2$. In order to understand this behavior, we need to additionally consider the orbital hybridization between the Bi square net and its neighboring Sr atoms. 
Along the $\Gamma$-M line, the linear band dispersion is dominated by the overlap of Bi ${p_{x,y}}$ orbitals, while it is dominated by the hybridization between Sr(Ca) \textit{d}-orbitals and Bi ${p_{x,y}}$ orbitals along the line normal to $\Gamma$-M.  Therefore, the stacking configuration of the akaline earth atoms that neighbor the square-net plays a decisive role in determining the band topology. A coincident akaline atom stacking leads to the Dirac band topology while a staggered ordering leads to the nodal-line band topology. Note that the discussion above does not consider SOC. When SOC is included, a small gap is introduced at the band-crossing point for both SrMnBi$_2$ and CaMnBi$_2$ \cite{lee2013anisotropic}.

Experimentally, the anisotropy of the Dirac fermions in SrMnBi$_2$ has been directly visualized by ARPES \cite{park2011anisotropic,jia2014observation,feng2014strong}.  A Fermi velocity anisotropy of \textit{(v$_{F,max}$/v$_{F,min}$)} $\geq$5 was extracted, in agreement with the theoretical prediction. For further confirmation of the Dirac nature of the fermions in SrMnBi$_2$, various magnetotransport \cite{park2011anisotropic,wang2011quantum,wang2012large,he2012giant, jo2014valley}, and optical conductivity \cite{park2017electrodynamic} measurements were performed. 
Based on the strong anistropy nature of the quasi-two-dimensional Dirac cone, an effective magnetic valley control in valley-polarized interlayer current was also demonstrated \cite{jo2014valley}. 


The fact that magnetism and Dirac fermions coexist in these materials raises the question whether there is any observable coupling between magnetic order and the nontrivial band topology. So far there is no indication magnetism effectively manipulates the Dirac fermions in SrMnBi$_2$ and CaMnBi$_2$. Inelastic neutron scattering experiments performed on SrMnBi$_2$ and CaMnBi$_2$ found no indications that the magnetic dynamics are influenced by the Dirac fermions \cite{rahn2017spin}. The situation is different however in \textit{MXZ$_2$} compounds that contain a magnetic rare-earth element on the \textit{M} position. Based on the crystal structure and band hybridization picture, the rare-earth atoms should possess a more direct interaction with the Bi square net than Mn atoms. This is the case in EuMnBi$_2$, which is isostructural to SrMnBi$_2$ \cite{may2014effect}. 
The temperature dependent magnetization and specific heat suggest divalent Eu$^{2+}$, which orders antiferromagnetically below \textit{T}$_N$ = 22\,K. Resonant x-ray scattering measurements showed that the Eu$^{2+}$ moments order ferromagnetically within the \textit{ab} plane and stack in the sequence of up-up-down-down along the \textit{c}-axis \cite{masuda2016quantum}. Note that Eu$^{2+}$ magnetic moments order slightly different than that of Ce$^{3+}$ in CeSbTe. In EuMnBi$_2$, the magnetic moments of the two neighboring rare earth planes above and below the pnictogen square-net are antiparallelly aligned, while in CeSbTe, they are aligned in parallel. Nevertheless, both of them show metamagnetic transition when field is applied along the easy axis \textit{c}, it happens at a roughly 20 times higher field (\textit{H}$_f\approx$ 5.3\,T) for EuMnBi$_2$. The existence of such metamagnetic transition makes EuMnBi$_2$ a promising candidate to demonstrate the Dirac fermions engineering by magnetism. 
Magnetotransport studies with an applied magnetic field of up to 55\,T, suggest the existence of a multilayer quantum hall effect \cite{masuda2016quantum} in EuMnBi$_2$.
While $\rho_{zz}$ is almost independent of field above \textit{T}$_N$, it exhibits a large jump with superpositions of significant SdH oscillations in the spin-flop phase. On the other hand, the inverse of in-plane Hall resistivity 1/$\rho_{yx}$ shows clear plateaus at regular intervals of \textit{B}$_F$/\textit{B}, where \textit{B}$_F$ is the frequency of SdH oscillation. The positions of these plateaus reasonably match the quantum oscillation minima of the in-plane resistivity $\rho{xx}$. The appearance of a multilayer quantum hall effect in a bulk, rather than 2D material was attributed to suppressed interlayer charge hopping and well confined Dirac fermions within the Bi square net in the field-induced spin-flop phase.


In addition to the materials in the CaSmP$_2$ and HfCuSi$_2$ structures, materials adopting the SrMnSb2$_2$ structure type were also investigated. SrMnSb2$_2$,  CaMnSb$_2$ and SrZnSb$_2$ (\textbf {Table \ref{tab1e2}}) are three representative materials that have been suggested to exhibit Dirac band crossings. Structurally they are fairly closely related to the HfCuSi$_2$ structure, except that the otherwise ideal 4$^4$ square nets are slightly distorted and the Sb atoms form into zig-zag chains along one of the in-plane directions. Therefore, the two orthogonal in-plane directions are anisotropic and the 4-fold rotation symmetry is lifted. We would expect that the band structure derived from the ideal 4$^4$ square net is affected by the distortion.
As we discussed earlier, the formation of zig-zag chains can gap the Dirac cones and result in fully gapped and trivial system. Whether that happens depends of the magnitude of the distortion. If the bond distance within the chains is much shorter than between the chains, it can be assumed that all electrons are localized within 2e2c bonds. If the difference in bond distance is less extreme, however, parts of the Dirac states can be retained. This is the case in SrMnSb2$_2$,  CaMnSb$_2$ and SrZnSb$_2$. 
Experimentally Sr$_{1-y}$Mn$_{1-z}$Sb$_2$ (with \textit{y},\textit{z} $<$ 0.1), was reported to possess a canted antiferromagnetic order in the Mn sublattice in the ground state, similar to that in YbMnBi$_2$. Therefore, Sr$_{1-y}$Mn$_{1-z}$Sb$_2$ provides another platform to explore the effect of time-reversal-symmetry breaking on Dirac bands.

There are quite a few further MXZ$_2$ compounds (see \textbf {Table \ref{tab1e2}}) that are reported to have nontrivial band topology. For detailed descriptions of their properties, readers are encouraged to read the references given in \textbf{Table \ref{tab1e2}}. We note that many of these compounds were reported to be topological materials based on the results of quantum oscillations. However, as already cautioned by \textit{Ando} \cite{ando2013topological}, the evaluation of the Berry phase in the quantum oscillations requires extra care and needs to be performed in an appropriate way, which is why many of these results remain ambiguous. 

\section{SUMMARY AND OUTLOOK}
We have reviewed the basic electronic structure of a square net of atoms and why it facilitates linear band-crossings. We provided a structural map linking the structural details of most known PbFCl-type materials to their electronic properties. The map indicates that compounds in which the square net is separated from other structural features (i.e. they are more 2D) are likely good candidates for TSMs. This is easy to understand, since in this case the electronic structure should be dominated by the square net. Thus simple geometric considerations help to ``separate the wheat from the chaff" within the large family of square-net materials. In combination with electron counting, clean TSMs can be identified.
Finally we have provided an overview of the electronic structure and physical properties reported in square-net based compounds. Many exotic properties, ranging from high mobility, large unsaturated magnetoresistance, and low effective mass, to more exotic electron-hole tunneling or bulk half-integer quantum Hall behavior, have been reported in this family. This large variety of properties highlights the rich physics and functionality potential that this material class bears.
 
The field of topological square-net materials has just begun and holds promise for new discoveries. New properties are still being explored and the large number of square-net-based compounds known provides the possibility for tuning the desired properties. But the library of topological square-net compounds is also not yet finalized. The periodic table in \textbf{Figure \ref{PSE}} contains some gaps: It is likely that the family of synthesizable square-net materials is not yet complete. Structural maps with more sophisticated or comprehensive considerations could be created for remaining square-net materials. This would help identifying the missing members of the family of square-net TSMs immensely.



\section*{DISCLOSURE STATEMENT}
 The authors are not aware of any affiliations, memberships, funding, or financial holdings that
might be perceived as affecting the objectivity of this review. 

\section*{ACKNOWLEDGMENTS}
This research was partially supported by NSF through the Princeton Center for Complex Materials, a Materials Research Science and Engineering Center DMR-1420541, and by a MURI grant on Topological Insulators from the Army Research Office, grant number ARO W911NF-12-1-0461. We would like to thank Raquel Queiroz for providing the tight-binding model that was used to create Figure \ref{sqbands}. LMS would like to thank J\"urgen Nuss and Andreas Topp for helpful discussions.

%


\noindent




\end{document}